\documentclass[aip,pof
,preprint
]{revtex4-1}

\usepackage{amsmath}

\usepackage{xcolor}

\usepackage{hyperref}

\usepackage{graphicx}% Include figure files
\usepackage{dcolumn}% Align table columns on decimal point
\usepackage{bm}% bold math

\begin{document}

\title{Two-Point Vorticity Statistics in the Inverse Cascade of Two-Dimensional Turbulence}

\author{R. Friedrich}
\affiliation{Institute for Theoretical Physics, University of M\"unster, Wilhelm-Klemm-Str.~9, D-48149 M\"unster, Germany}
\author{M. Vo{\ss}kuhle}
\affiliation{Laboratoire de Physique, ENS de Lyon, 46 all\'{e}e d'Italie F-69007 Lyon, France}
\author{O. Kamps}
\email{okamp@uni-muenster.de}
\affiliation{Center for Nonlinear Science, University of M\"unster, Corrensstr. 2, D-48149 M\"unster, Germany}
\author{M. Wilczek}
\affiliation{Institute for Theoretical Physics, University of M\"unster, Wilhelm-Klemm-Str.~9, D-48149 M\"unster, Germany}

\date{\today}

\begin{abstract}
A statistical analysis of the two-point vorticity statistics in the inverse energy cascade of two-dimensional turbulence is presented in terms of probability density functions (PDFs). Evolution equations for the PDFs are derived in the framework of the Lundgren--Monin--Novikov hierarchy, and the unclosed terms are studied with the help of direct numerical simulations (DNS). Furthermore, the unclosed terms are evaluated in a Gaussian Approximation and compared to the DNS results. It turns out that the statistical equations can be interpreted in terms of 
the dynamics of screened vortices. The two-point statistics is related to the
dynamics of two point vortices with screened velocity field, where an effective relative motion of the two point vortices originiating from the turbulent surroundins is identified to be a major characteristics of the dynamics
underlying the inverse cascade.
\end{abstract}


\maketitle

\section{Introduction}

Fully developed turbulence is an example for a system far from equilibrium giving rise to a transport of energy and enstrophy across scales. In fact, the existence of these fluxes can be seen as a distinguishing feature of turbulence. While the famous four-fifth law for three-dimensional turbulence proves the existence of a direct energy transfer across scales, the inclusion of this fact into a complete statistical theory derived from the Navier--Stokes equation remains one of the main challenges \cite{Monin2007B,Tsinober2009}.

While three-dimensional turbulence is characterized by an energy and enstrophy flux towards smaller scales, two-dimensional turbulent flows play a peculiar role due to the possibility of a direct as well as an inverse cascade, as has been emphasized in the seminal work of Kraichnan \cite{kraichnan67phf}. The coexistence of both cascades has been convincingly demonstrated by Boffetta and Musacchio \cite{boffetta10pre}. Recently, a detailed analysis of the contour lines of zero vorticity has revealed interesting scaling behavior pointing to the existence of nontrivial multi-point statistics of vorticity in the inverse cascade \cite{bernard06nph}. Especially the inverse cascade is an interesting case to study, because the fact that energy is transferred to larger scales, which already contain the major amount of energy, seems counter-intuitive \cite{Cardy2009}. Thus an explanation of this fact can be regarded as a challenge for classical non-equilibrium physics. 

A central goal of the present work is to contribute to a deeper understanding of the inverse energy cascade in terms of the vorticity. To this end, we study the two-point vorticity statistics as well as the vorticity increment statistics in the framework of the Lundgren--Monin--Novikov (LMN) hierarchy \cite{lundgren67pf,monin67jamm,novikov68spd}. The unclosed terms will be estimated from direct numerical simulations. It turns out that the results can be intuitively interpreted by the method of characteristics, which naturally leads to a semi-Lagrangian description of the statistical evolution across scales. These observations are paralleled by calculations of the unclosed terms from a Gaussian Approximation (GA), which already reproduces a number of observed features. The inverse energy cascade, however, turns out not to be captured within these calculations revealing the non-perturbative nature of the problem. A comparison to the DNS results then shows which features have to be incorporated into a comprehensive theory.

The remainder of this article is structured as follows. After introducing the basic physical setting of two-dimensional turbulence in the inverse cascade regime in Section~\ref{sec:equations}, we establish the statistical framework in Section~\ref{sec:statistical_treatment}. In Section~\ref{sec:lga} an approximate calculation of the unclosed terms in case of the two-point vorticity statistics along with a comparison to DNS data is presented. We then proceed to a detailed examination of the vorticity increment statistics in Section~\ref{sec:vorticity_increments}, before we conclude.

\begin{figure}
\centering
\includegraphics{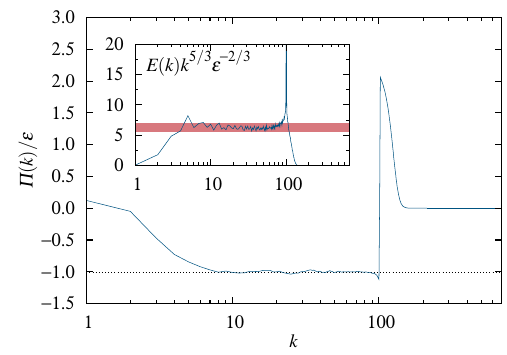}
\caption{\label{fluxfig}Spectral energy flux obtained from simulation A. The compensated spectrum exhibits a clear indication for the existence of the inverse energy cascade. The red line in the inset marks the band of Kolmogorov constants reported in earlier publications \cite{boffetta00pre,goto04njp,paret97prl}.}
\end{figure}

\section{\label{sec:equations}Equations of Motion, Numerics, and Basic Results}

\begin{table}\caption{\label{tab:parameters}Parameters for simulations A and B. Specified are the timestep $\Delta t$, the viscosity and the friction constant, $\nu_H$ and $\nu_V$ respectively, the forcing wavenumber $k_f$, the integral length scale $L_\mathrm{int}$, the Taylor length $\lambda$, and the large-eddy turnover time $T$.
}
\begin{ruledtabular}
\begin{tabular}{lD..{6}D..{8}D..{1}D..{0}D..{1}D..{3}D..{3}D..{3}}
 & \multicolumn{1}{c}{$\Delta t$} & \multicolumn{1}{c}{$\nu_H$} & \multicolumn{1}{c}{$\nu_V$} & \multicolumn{1}{c}{$k_f$} & \multicolumn{1}{c}{$k_b$} & \multicolumn{1}{c}{$L_\mathrm{int}$} & \multicolumn{1}{c}{$\lambda$} & \multicolumn{1}{c}{$T$} \\
\hline
Simulation~A & 0.000125 & 2.5\times 10^{-32} & 2.5 & 102 & 1.5 & 0.375 & 0.066 & 0.089\\
Simulation~B & 0.000125 & 2.0\times 10^{-41} & 2.5 & 411 & 1.5 & 0.316 & 0.027 & 0.073\\
\end{tabular}
\end{ruledtabular}
\end{table}

The spatio-temporal dynamics of the turbulent field is governed by the two-dimensional vorticity equation
\begin{equation}
\frac{\partial }{\partial t} \omega(\bm{x},t)+\bm{u}(\bm{x},t)\cdot \nabla_{\bm{x}} \omega(\bm{x},t)=L(-\Delta_{\bm{x}}) \omega(\bm{x},t)+F(\bm{x},t) \, .
\end{equation}
The velocity field $\bm{u}(\bm{x},t)$ is determined from vorticity
$\omega(\bm{x},t)$ via Biot--Savart's law,
\begin{equation} \label{eq:BiotSavart}
\bm{u}(\bm{x},t)=\int \! \mathrm{d} \bm{x}'  \bm{e}_z \times
\frac{\bm{x}-\bm{x}'}{2\pi |\bm{x}-\bm{x}'|^2} \omega(\bm{x}',t) \, 
\end{equation}
Dissipation is taken into account by the operator $L(-\Delta_{\bm{x}})= f_V(-\Delta_{\bm x})+f_H(-\Delta_{\bm{x}})$, where we allow for the presence of hyperviscosity as well as friction acting on the large scales. In order to achieve a constant energy flux in the inertial range (see Fig.\,\ref{fluxfig}) we have used a hyperviscosity of the form $f_H(-\Delta_{\bm{x}})= \nu_H (-\Delta_{\bm{x}})^8$ and a large scale friction $f_V(-\Delta_{\bm x})=-\nu_V(-\Delta_{\bm{x}})^{-1}$ to extract energy transported to the large scales by the inverse cascade. In contrast to the normal viscous term the hyperviscous term is a Laplacian to the power of eight. This makes it necessary to choose $\nu_H$ very small (see Table~\ref{tab:parameters}), so that the dissipation becomes only relevant for wavenumbers that are bigger than the forcing wavenumber. Energy is injected into the system by the forcing term $F(\bm{x},t)$. Here we employ a forcing that acts only on a narrow wavenumber band of width $k_b$ around $k_f$ in Fourier space. Within this band the amplitudes $|\omega_{\bm{k}}|$ of the Fourier modes $ \omega_{\bm{k}}(t)=|\omega_{\bm{k}}|\mathrm{e}^{i \Phi_{\bm{k}}(t)} $ are kept constant by resetting them every time step to the constant $|\omega_{\bm{k}}|= \mathrm{const.}$ with $k_f-k_b /2<|\bm{k}|< k_f+ k_b /2$ while the phases can evolve freely according to the vorticity equation \cite{goto04njp}. 

The vorticity equation is numerically solved by means of a standard pseudospectral method with $2/3$-dealiasing in a periodic box with side length $2\pi$ and a resolution of $2048^2$ grid points. Time stepping is performed by a memory saving Runge--Kutta scheme of third order \cite{shu88jcp}. The hyperviscous term is treated by an integrating factor. Data-analysis has been performed over more than 200 snapshots of the vorticity field taken from a simulation in a statistically stationary state. The snapshots are equally separated by about 1.4 large-eddy turnover times.

In most simulations (see e.g. Refs.\,\onlinecite{boffetta00pre, goto04njp,paret97prl}) the forcing is located at a wavenumber $k_f$ that is close to the highest wavenumber resolved. In combination with hyperviscosity this allows for a wide inertial range and suppression of the direct enstrophy cascade. However, this leads to a poor numerical  resolution of the forcing scale. To circumvent this problem we inject the energy at an intermediate length scale $k_f = 102$ and use a hyperviscosity that, for numerical reasons, saturates at a maximal value \footnote{We define a maximal wavenumber $k_H^\mathrm{max} = 170$ such that the viscous term does not grow beyond this scale, i.e. $f_H(-\Delta_{\bm{x}}) \leq \nu_H [ ({k_H^\mathrm{max}})^{2}]^{8}$.} (this simulation is denoted as simulation A). The viscosity was chosen as $\nu_H = 2.5 \times 10^{-32}$. Comparison to a reference simulation (simulation B) with $k_f= 411$ and $\nu_H = 2.0 \times 10^{-41}$ showed that this procedure does not affect the results while yielding a better resolution at the forcing scale. Simulation A has a Taylor length scale of $\lambda= 0.066$ and an integral length scale of $L=0.375$. For simulation B we get $\lambda= 0.027$ and $L = 0.316$. In both simulations we have $\nu_V = 2.5$. These values are summarized in Table~\ref{tab:parameters}.

\begin{figure}
\centering
\includegraphics{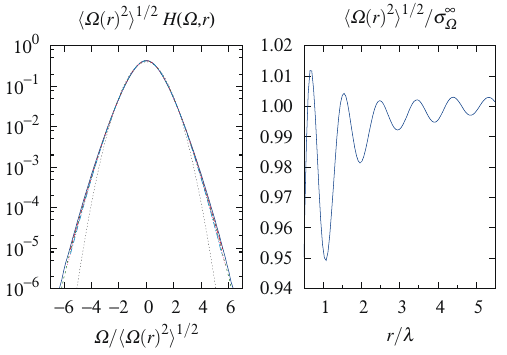}
\caption{\label{pdf}Left: Probability distribution of the standardized vorticity increment $\Omega(r,t)/\sqrt{\langle \Omega(r,t)^2 \rangle}$ as a function of $r$ ($r=\{1,2,\dots, 5\}\lambda$). The Taylor scale $\lambda$ is defined by $[2\langle u^2 \rangle/\langle (\partial_x u)^2\rangle]^{1/2}$. Right: Second moment $\langle \Omega(r,t)^2 \rangle$ of the vorticity increment as a function of scale $r$. The moment exhibits slight oscillations in scale. The constant $\sigma_\Omega^\infty$ denotes the value of the second moment in the limit $r\rightarrow \infty$, i.e. $\sigma_\Omega^\infty = (2\langle\omega^2\rangle)^{1/2}$. Both results were obtained from simulation A.}
\end{figure}

As a characterization of the simulation A, Fig.\,\ref{fluxfig} shows the energy spectrum in Fourier-space in the inverse cascade, as obtained from our numerical solution. The spectral energy flux is constant and demonstrates the existence of the inverse cascade. The obtained Kolmogorov constant $K_o$, defined by $E(k)=K_o\epsilon^{2/3}k^{-5/3}$ is consistent with the results reported in the literature \cite{boffetta00pre,goto04njp,paret97prl}. The PDFs of the vorticity increments, defined according to $\Omega(\bm{x},\bm{r},t)=\omega(\bm{x}+\bm{r},t)-\omega(\bm{x},t)$, are shown in Fig.\,\ref{pdf} as a function of scale $r=|\bm{r}|$. Apparently, the vorticity increment is nearly Gaussian distributed with slight super-Gaussian tails. Furthermore, the second moment of the vorticity increment exhibits oscillations as a function of the distance $r$, which are clearly a result of the small but visible oscillations in the shape of the PDFs. In the following we shall be concerned with the behavior represented by the center of the PDF, where a Gaussian representation of the statistics is possible. The behavior of the tails of the PDF may be accessible by instanton calculations \cite{Falkovich2011pre,Moriconi2004pre}.

Fig.\,\ref{pdf} and Fig.\,\ref{grfig} show that the second-order moment of vorticity increments and the vorticity-vorticity correlation function, respectively, are oscillating on the scale of the Taylor length. This gives us some information on the formation of the inverse cascade. The small-scale forcing which is confined to a narrow band of wavenumbers acts directly on the vorticity field and has an oscillating autocorrelation function. Physically this kind of forcing creates a field of small vortices where it is more likely to find small vortices with opposite signs next to each other. The size of the vortices is directly related to the forcing length scale. The interesting fact is that the formation of the inverse cascade preserves this small scale structure in a way that these oscillations are also observable directly in the vorticity field. This indicates that, as has been stressed by Paret and Tabeling~\cite{paret97prl}, the inverse cascade is not based on vortex merging, but on a rearrangement or aggregation of small vorticity patches, which, without permanent forcing, would lead to large clusters of vorticity and, eventually, to the condensation of system size vortices. The large-scale vortices, however, are destroyed by small-scale forcing, i.e. by a change of the circulation of the small scale vortices. However, the tendency of aggregation of like-signed vortices characterizes the system's behavior. This means that the formation of the inverse cascade is a collective effect of interacting small vortices rather than the formation of large-scale vorticity structures. A similar behavior was observed in \cite{boffetta00pre} where the small vortices have been created by a different type of forcing.

\section{\label{sec:statistical_treatment}Statistical Treatment}
Next, we proceed to a statistical description of two-dimensional turbulence, which is based on the $N$-point vorticity PDF $f(\lbrace \omega_l,{\bm{x}}_l \rbrace,t)$. This PDF can be defined as an ensemble average over the fine-grained probability distribution, $f(\lbrace \omega_l,{\bm{x}}_l \rbrace,t)=\langle \tilde f(\lbrace \omega_l,{\bm{x}}_l \rbrace,t) \rangle$, where the fine-grained PDF is defined according to:
\begin{equation}
\tilde f(\lbrace \omega_l,{\bm{x}}_l \rbrace, t) = \prod_{l=1}^N \delta(\omega_l
-\omega({\bm{x}}_l,t))
\end{equation}
Here, $\omega({\bm{x}}_l,t)$ denotes the solution of the vorticity equation at spatial point ${\bm{x}}_l$ and time $t$. It is straightforward to derive the following evolution equation for the fine-grained PDF \cite{friedrich10cp}; a short discussion of the procedure can be found in Appendix~\ref{app:pdf}. The resulting equation takes the form
\begin{multline}
\frac{\partial }{\partial t} \tilde f(\lbrace \omega_l,{\bm{x}}_l \rbrace,t) +
\sum_{i=1}^N \nabla_{{\bm{x}}_i} \cdot \big[\bm{u}({\bm{x}}_i,t) \tilde f(\lbrace
\omega_l,{\bm{x}}_l \rbrace,t)\big] \\
=  -\sum_{i=1}^N \frac{\partial }{\partial \omega_i} \big[
L(-\Delta_{{\bm{x}}_i}) \omega({\bm{x}}_i,t)+ F({\bm{x}}_i,t) \big]  \tilde f(\lbrace
\omega_l,{\bm{x}}_l \rbrace,t) \, .
\end{multline}
In order to obtain the evolution equation for the ensemble PDF $f(\lbrace \omega_l,{\bm{x}}_l \rbrace,t)$, one has to perform an ensemble average of this equation. This averaging procedure immediately confronts us with the closure problem of turbulence, since the following unclosed expectations arise:
\begin{eqnarray}\label{cond1}
\langle \bm{u}(\bm{x},t) \tilde f(\lbrace \omega_l,{\bm{x}}_l \rbrace,t) \rangle &=&
\langle \bm{u}(\bm{x},t)|\lbrace \omega_l,{\bm{x}}_l\rbrace \rangle f(\lbrace
\omega_l,{\bm{x}}_l \rbrace,t)  \nonumber \\
\langle L(-\Delta_{\bm{x}})\bm{u}(\bm{x},t) \tilde f(\lbrace \omega_l,{\bm{x}}_l
\rbrace,t) \rangle
 &=& \langle L(-\Delta_{\bm{x}}) \omega(\bm{x},t)|\lbrace \omega_l,{\bm{x}}_l\rbrace
\rangle f(\lbrace \omega_l,{\bm{x}}_l \rbrace,t)
\nonumber \\
\langle F(\bm{x},t) \tilde f(\lbrace \omega_l,{\bm{x}}_l \rbrace,t)\rangle &=& \langle
F(\bm{x},t)|\lbrace \omega_l,{\bm{x}}_l\rbrace \rangle f(\lbrace \omega_l,{\bm{x}}_l
\rbrace,t)
\end{eqnarray}
It is convenient to express these unclosed expectations, as indicated above, in  terms of conditional averages of velocity, $\langle \bm{u}(\bm{x},t)|\lbrace \omega_l,{\bm{x}}_l \rbrace \rangle$, diffusion $\langle L(-\Delta_{\bm{x}}) \omega (\bm{x},t)|\lbrace \omega_l,{\bm{x}}_l\rbrace \rangle$, which is related to the dissipation of enstrophy, and forcing $\langle F(\bm{x},t)|\lbrace \omega_l,{\bm{x}}_l\rbrace \rangle $. The introduction of these quantities is motivated by the fact that conditional averages can be estimated from DNS yielding deeper insights in the shape and evolution of PDFs of various turbulent quantities \cite{novikov94mplb,mui96pre,wilczek09pre,Wilczek2011jfm,luelff11njp}.

An alternative way is to write down the whole chain of evolution equations for higher $N$-point functions, a procedure which is well documented in the literature \cite{Monin2007B,ulinich69jetp,lundgren67pf,novikov1965A,novikov68spd}. However, up to now conclusive truncations of the hierarchy are lacking. Additionally, we want to mention that a similar approach for the study of the velocity increment statistics of the inverse cascade has been undertaken by Boffetta et al. \cite{boffetta02pre} based on theoretical approaches by V.\,Yakhot \cite{yakhot01pre,yakhot99pre}. Here, numerical input for the conditional pressure term is used to close the equation for the PDF of the velocity increment.

Using the conditional expectations~\eqref{cond1} we arrive at the following partial differential equation determining the PDF:
\begin{align}\label{evolutionPDF}
 \frac{\partial }{\partial t} f(\lbrace \omega_l,{\bm{x}}_l \rbrace,t ) + \sum_{i=1}^N
\nabla_{{\bm{x}}_i} \cdot \big[\langle \bm{u}({\bm{x}}_i,t)|\lbrace \omega_l,{\bm{x}}_l
\rbrace \rangle & f(\lbrace \omega_l,{\bm{x}}_l \rbrace,t)\big] =  \nonumber \\ -\sum_{i=1}^N
\frac{\partial }{\partial \omega_i} \big[ \langle L(-\Delta_{{\bm{x}}_i}) \omega({\bf
x}_i,t)|\lbrace \omega_l,{\bm{x}}_l\rbrace \rangle
&+  \langle F({\bm{x}}_i,t)|\lbrace \omega_l,{\bm{x}}_l \rbrace \rangle \big]
f(\lbrace \omega_l,{\bm{x}}_l \rbrace,t) \, .
\end{align}
These partial differential equations can be solved by the method of characteristics\cite{Courant1989}. The corresponding characteristic equations read
\begin{align}\label{charac1}
\dot {\bm{x}}_i &= \langle \bm{u}({\bm{x}}_i,t)|\lbrace \omega_l,{\bm{x}}_l \rbrace \rangle 
 \\
\dot \omega_i &= \label{charac2} 
\langle L(-\Delta_{{\bm{x}}_i}) \omega({\bm{x}}_i,t)|\lbrace \omega_l,{\bm{x}}_l\rbrace 
\rangle 
+ \langle F({\bm{x}}_i,t)|\lbrace \omega_l,{\bm{x}}_l\rbrace \rangle \, .
\end{align}
Along these characteristic curves the probability distribution changes according to
\begin{align}
\frac{\mathrm{d} }{\mathrm{d} t} f(\lbrace \omega_l,{\bm{x}}_l \rbrace,t) &+\sum_i \bigg[\nabla_{{\bm{x}}_i}\cdot \langle \bm{u}({\bm{x}}_i,t)|\lbrace \omega_l,{\bm{x}}_l \rbrace \rangle \bigg] f(\lbrace \omega_l,{\bm{x}}_l \rbrace,t) \nonumber \\
&+\sum_i \left[
\frac{\partial }{\partial \omega_i}
\langle L(-\Delta_{{\bm{x}}_i}) \omega({\bm{x}}_i,t)|\lbrace \omega_l,{\bm{x}}_l\rbrace
\rangle \right]
f(\lbrace \omega_l,{\bm{x}}_l \rbrace,t)
\nonumber \\
&+\sum_i \left[
\frac{\partial }{\partial \omega_i}
\langle F({\bm{x}}_i,t)|\lbrace \omega_l,{\bm{x}}_l\rbrace \rangle \right]
f(\lbrace \omega_l,{\bm{x}}_l \rbrace,t) \nonumber \\
&= 0 \, .
\end{align}
The characteristics still contain the unclosed terms, and therefore it is not possible to solve this set of ordinary differential equations analytically. Nevertheless, this kind of approach is extremely valuable since it leads us to an interpretation of the dynamics of the $N$-point vorticity statistics in terms of point vortex-like objects. Equation~\eqref{charac1} can be intuitively interpreted in the way that the vorticity at position ${\bm{x}}_i$ is advected by the conditional velocity field at this position. Equation~\eqref{charac2} describes the change of vorticity along the trajectory due to the dissipative terms and the forcing. 

The numerical data presented in the current work are generated with a deterministic forcing term, but it is also quite common to use a stochastic forcing term. In this case, the Navier--Stokes equation turns into a stochastic partial differential equation. The stochastic forcing term considered is specified by \cite{Risken96}   
\begin{equation}\label{whiteNoise}
\langle F({\bm{x}}_i,t)F({\bm{x}}_j,t') \rangle=2 Q({\bm{x}}_i-{\bm{x}}_j) \delta(t-t') \, .
\end{equation}
Under this assumption the conditional expectation for the forcing term $\langle F({\bm{x}}_i,t)|\lbrace \omega_i,{\bm{x}}_i\rbrace \rangle$ in Eq.\,\eqref{evolutionPDF} appears as a second-derivative term of the form \cite{novikov1965A}
\begin{equation}
 -\frac{\partial}{\partial \omega_i} \langle F({\bm{x}}_i,t)|\lbrace \omega_l,{\bm{x}}_l \rbrace \rangle f(\lbrace \omega_l,{\bf
x}_l \rbrace,t)= \sum_{ij}  Q({\bm{x}}_i-{\bm{x}}_j) \frac{\partial }{\partial
\omega_i}\frac{\partial }{\partial \omega_j} f(\lbrace
\omega_l,{\bm{x}}_l \rbrace,t) \, .
\end{equation}
As a result the PDF equation turns into a second-order partial differential equation. Consequently the characteristic equation for the vorticity~\eqref{charac2} turns into a stochastic equation with the stochastic forcing $F_i(t)$
\begin{equation}
 \dot \omega_i =  \langle L(-\Delta_{{\bm{x}}_i}) \omega({\bm{x}}_i,t)|\lbrace \omega_l,{\bf
x}_l\rbrace \rangle + F_i(t) \, .
\end{equation}

\section{\label{sec:lga}Gaussian Approximation and DNS Results}

Up to now, the statistical equations are exact, but unclosed. To proceed, we either have to use DNS data to evaluate the unclosed terms or rely on a closure approximation to evaluate the unclosed terms. Both approaches will be followed in this section.

\subsection{Gaussian Approximation for N points}

Since the two-point statistics of vorticity is close to Gaussian, it is tempting to approximate the conditional averages~\eqref{cond1} based on the assumption of Gaussian statistics. It turns out that the statistical quantities arising in our framework depend on the first conditional moment of the vorticity field with respect to fixed sample-space vorticities. In the GA, this conditional average is a linear function of these sample-space vorticities. The details of this calculation are given in Appendix~\ref{app:lga}.

The conditional velocity field can be expressed via Biot--Savart's law in terms of the conditional vorticity expectation $\langle \omega(\bm{x}',t)|\lbrace \omega_l,{\bm{x}}_l\rbrace \rangle$ (with $l=1,\dots,N$) according to 
\begin{equation}
\langle \bm{u}(\bm{x},t)|\lbrace \omega_l,{\bm{x}}_l \rbrace \rangle = \int \! \mathrm{d}
\bm{x}' \bm{e}_z \times \frac{\bm{x}-\bm{x}'}{2\pi|\bm{x}-\bm{x}'|^2} \langle
\omega(\bm{x}',t)|\lbrace \omega_l,{\bm{x}}_l \rbrace \rangle \, .
\end{equation}
In the case of Gaussian statistics all quantities depend only on the vorticity-vorticity correlation function $C(\bm{x}'-\bm{x})=\langle \omega(\bm{x},t) \, \omega(\bm{x}',t) \rangle/\langle \omega(\bm{x},t)^2 \rangle$. The conditional expectation is then given by \cite{friedrich10cp} (see also Appendix~\ref{app:lga} of this paper)
\begin{equation}\label{condomega}
\langle \omega(\bm{x},t)|\lbrace \omega_l,{\bm{x}}_l \rbrace \rangle=\sum_{km} C({\bf
x}-{\bm{x}}_k)C({\bm{x}}_k-{\bm{x}}_m)^{-1}\omega_m
\end{equation}
so that we obtain the following linear approximations for conditional velocity
\begin{equation}\label{condvel}
\langle \bm{u}(\bm{x},t)|\lbrace \omega_l , {\bm{x}}_l \rbrace \rangle = \sum_{km} {\bf
U}(\bm{x}-{\bm{x}}_k) C({\bm{x}}_k-{\bm{x}}_m)^{-1} \omega_m \, .
\end{equation}
Here, $C({\bm{x}}_k-{\bm{x}}_m)^{-1}$ denotes the ($k$,$m$)-th element of the inverse of the matrix defined by two-point correlation functions of all vorticities under consideration (see Appendix \ref{app:lga} for further details). Furthermore, we have introduced the velocity kernel
\begin{eqnarray}\label{dressed}
\bm{U}(\bm{x}-{\bm{x}}_i) &=&
\int \! \mathrm{d}\bm{x}' \, \bm{e}_z \times \frac{\bm{x}-\bm{x}'}{2\pi|\bm{x}-\bm{x}'|^2}
C(\bm{x}'- {\bm{x}}_i) \, .
\end{eqnarray}
The field  $\bm{U}(\bm{x}-{\bm{x}}_i)$ is the velocity field of a vorticity distribution $C(\bm{x}-{\bm{x}}_i)$ centered at ${\bm{x}}_i$. A point vortex with circulation $\Gamma_i$ would be described by $C(\bm{x}-{\bm{x}}_i)=\Gamma_i \delta(\bm{x}-{\bm{x}}_i)$. As a consequence we can visualize the velocity field $\bm{U}(\bm{x}-{\bm{x}}_i)$ as the field of a {\em screened vortex}, i.e. a quasi-vortex in the sense of a Landau quasi-particle taking into account the surrounding turbulence by an {\em effective velocity} profile. However, one has to keep in mind that these quasi-vortices are strongly interacting via a modified vortex dynamics, in contrast to the weakly interacting Landau quasi-particles. The complete conditional velocity field is a linear superposition of velocity fields of such screened (or {\em dressed}) vortices. Alternatively, one could also say that Biot--Savart's law is changed by the effective consideration of the turbulent background field. We mention that a change of Biot--Savart's law has been used by Chevillard et al. \cite{chevillard10epl} in the modeling of the statistics of velocity increments in three-dimensional turbulence.

To proceed, the conditional vorticity diffusion (related to the enstrophy dissipation) is obtained in the GA as
\begin{equation}
\langle L(-\Delta_{{\bm{x}}})\omega(\bm{x},t)|\lbrace \omega_l , {\bm{x}}_l \rbrace
\rangle =
-\sum_{km} \gamma(\bm{x}-{\bm{x}}_k) C({\bm{x}}_k-{\bm{x}}_m)^{-1} \omega_m \, .
\end{equation}
Here, the dissipation field $\gamma(\bm{x}-{\bm{x}}_i)$ is defined according to
\begin{equation}\label{conddiss}
\gamma(\bm{x}-{\bm{x}}_i)=-L(-\Delta_{\bm{x}}) C(\bm{x}-{\bm{x}}_i) \, .
\end{equation}
The characteristic equations then take the form
\begin{align}\label{chaga}
\dot {\bm{x}}_i &= \sum_{km}
\bm{U}({\bm{x}}_i-{\bm{x}}_k)C({\bm{x}}_k-{\bm{x}}_m)^{-1}
\omega_m \nonumber \\
\dot \omega_i &= -\sum_{km} \gamma({\bm{x}}_i-{\bm{x}}_k)
C({\bm{x}}_k-{\bm{x}}_m)^{-1}\omega_m+F({\bm{x}}_i,t) \, .
\end{align}
This explicitly shows how the notion of quasi-particles enters the theoretical description in a natural way, as the characteristic equations~\eqref{chaga} can be interpreted in a Lagrangian sense. To render this particularly simple, we first neglect the temporal variation of the vorticities $\omega_i$. Then the set of differential equations for the positions ${\bm{x}}_i$ can be viewed as a modified point vortex system, where the velocity field of an individual vortex is not the one of a point vortex, but is the velocity field of a dressed vortex with a profile modified by the two-point vorticity correlation. Letting in addition the amplitudes of vorticity vary due to the action of dissipation and external energy input adds further complexity to the dynamics. Hence Eqs.~\eqref{chaga} describe a set of modified point vortices with time-dependent amplitudes. This perspective can be taken in close analogy to the notion of a Landau quasi-particle, a notion which is widely used in theories  of many-particle systems. In these theories, part of the many-particle interaction can be absorbed into effective properties of single particles, so-called quasi-particles. In the present case, the reduction from a continuum description of vorticity to the description of the statistics of the field at finite spatial points quite naturally leads to an effective point vortex dynamics. We want to point out that in contrast to the Landau quasi-particles of condensed matter physics, the quasi-vortices are still strongly interacting. However, since we are still in the realm of the linear GA, we are not able to obtain the inverse cascade on the basis of a two-point closure due to the vanishing transport terms discussed below.

\begin{figure}
\centering
\includegraphics{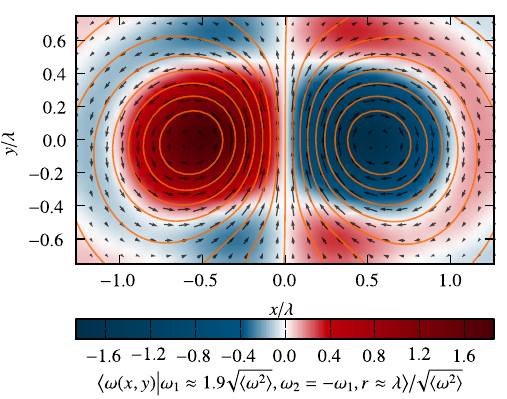}
\caption{\label{condv}Conditional vorticity and velocity fields, $\langle \omega(\bm{x},t) \vert \omega_1,{\bm{x}}_1;\omega_2,{\bm{x}}_2\rangle$ and $\langle \bm{u}(\bm{x},t) \vert\allowbreak \omega_1,{\bm{x}}_1;\allowbreak \omega_2,{\bm{x}}_2\rangle$ respectively, for the case $\omega_1 = -\omega_2 \approx 1.9 \langle\omega^2\rangle^{1/2}$ and $r=\lvert {\bm{x}}_2 - {\bm{x}}_1\rvert=\lambda$ obtained from DNS (simulation~A). The conditional vorticity is color-coded and the color-bar gives the corresponding values in multiples of the root mean square vorticity $\langle\omega^2\rangle^{1/2}$ . The arrows represent direction and magnitude of the conditional velocity at a subset of the grid points of our simulation. Finally, contour lines of the conditional stream function $\langle \Psi(\bm{x},t)\vert\omega_1,{\bm{x}}_1;\omega_2,{\bm{x}}_2\rangle$, where $\Psi(\bm{x},t) = -\Delta^{-1}\omega(\bm{x},t)$, are plotted in orange.
}
\end{figure}

\begin{figure}
\centering
\includegraphics{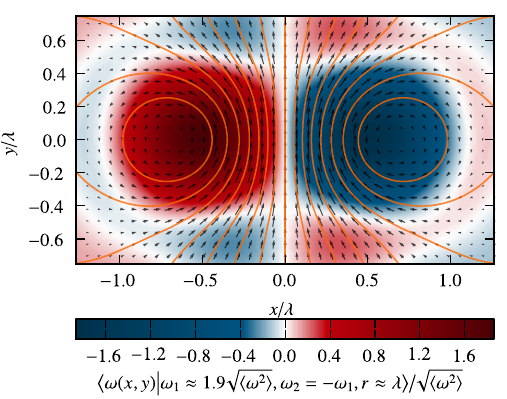}
\caption{\label{condgauss}Conditional vorticity and velocity fields as in Fig.\,\ref{condv}, but here evaluated following the GA, Eqs.\,\eqref{condomegaLGA} and\,\eqref{condvelLGA} respectively. The only quantity to be specified for this approximation is the correlation function which was estimated from the data of simulation~A.}
\end{figure}

\subsection{Gaussian Approximation for two points and comparison to DNS}

Up to now we have dealt with the GA for the case of the full $N$-point vorticity statistics. In order to compare the results of the GA with DNS we will restrict our considerations to the case of two spatial points, since then the estimation of  the conditional averages from DNS becomes feasible. To this end we numerically determined the conditional vorticity field $\langle \omega (\bm{x},t)\vert\omega_1,{\bm{x}}_1;\omega_2,{\bm{x}}_2\rangle$ and the conditional velocity field $\langle \bm{u}(\bm{x},t) \vert\omega_1,{\bm{x}}_1; \omega_2,{\bm{x}}_2\rangle$ and compared it with the ones obtained from the GA. Using~\eqref{condomega}, \eqref{condvel} and isotropy we find for the conditional vorticity field conditioned on vorticity $\omega_1$, $\omega_2$ at positions ${\bm{x}}_1$, ${\bm{x}}_2$ in this approximation
\begin{equation}\label{condomegaLGA}
\langle \omega (\bm{x},t)|\omega_1,{\bm{x}}_1;\omega_2,{\bm{x}}_2 \rangle
= C(|\bm{x}-{\bm{x}}_1|) \, G(\omega_1,\omega_2,|{\bm{x}}_1-{\bm{x}}_2|)
+ C(|\bm{x}-{\bm{x}}_2|) \, G(\omega_2,\omega_1,|{\bm{x}}_2-{\bm{x}}_1|)
\end{equation}
and for the conditional velocity field
\begin{equation}\label{condvelLGA}
\langle \bm{u}(\bm{x},t)|\omega_1,{\bm{x}}_1;\omega_2,{\bm{x}}_2 \rangle
= \bm{U}(|\bm{x}-{\bm{x}}_1|) \, G(\omega_1,\omega_2,|{\bm{x}}_1-{\bm{x}}_2|)
+ \bm{U}(|\bm{x}-{\bm{x}}_2|) \, G(\omega_2,\omega_1,|{\bm{x}}_2-{\bm{x}}_1|)
 \, .
\end{equation}
with
\begin{equation}
  G(\omega_1,\omega_2,|{\bm{x}}_1-{\bm{x}}_2|) = \frac{C(0) \omega_1-C(|{\bm{x}}_1-{\bf
    x}_2|)\omega_2}{C(0)^2-C(|{\bm{x}}_1-{\bm{x}}_2|)^2} \, .
\end{equation}
Furthermore, we note that $\bm{U}(0)=\bm{0}$, due to the fact that $C(0)$ is finite, c.f. Eq.\,\eqref{dressed}.

\begin{figure}
\centering
\includegraphics{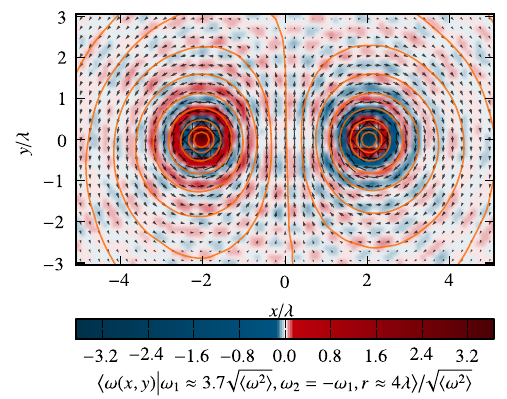}
\caption{\label{condvNew}Same quantities as in Fig.\,\ref{condv} now for $r\approx 4 \lambda$. To be able to show the conditional averages for large $r$ in terms of the Taylor scale, which is still deep in the inertial range, we have used simulation~B.}
\end{figure}

\begin{figure}
\centering
\includegraphics{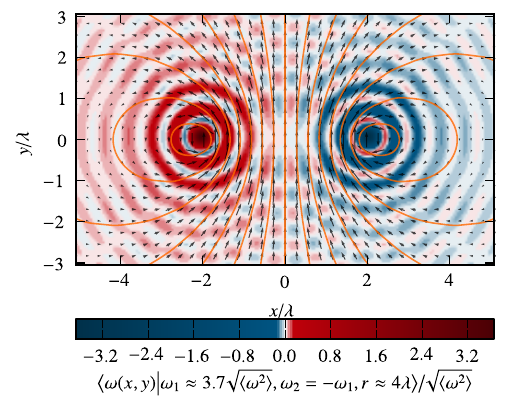}
\caption{\label{condgaussNew}GA as in Fig.\,\ref{condgauss} for $r\approx 4
\lambda$. The estimation of the correlation function was based upon simulation~B (as is Fig.\,\ref{condvNew}).}
\end{figure}

The visual impression of both fields is very similar, as can be seen from Figs.~\ref{condv}, \ref{condgauss} and~\ref{condvNew}, \ref{condgaussNew}, respectively. Here the conditional vorticity field with respect to two fixed vorticities is shown. For small separation, the conditional field reminds of the vorticity field generated by a vortex dipole. For larger distance, typical oscillations are visible. Also the velocity field induced by the conditional vorticity field is shown, which again strongly reminds of a vortex dipole. These features are captured excellently by the GA.

\subsection{The von~K\'{a}rm\'{a}n--Howarth equation for the vorticity and breakdown of the Gaussian Approximation}

In this section, we explicitly evaluate the analog of the von~K\'{a}rm\'{a}n--Howarth equation for the vorticity in the realm of the GA. As we shall see, the advective term related to the enstrophy flux across scales vanishes in this approximation, i.e., there is no enstrophy cascade. This also implies a vanishing energy transfer across scales. This breakdown of the GA is related to the small, but significant asymmetries that are observed in the DNS results, but which are not present in the GA.

The evolution equation for the vorticity covariance at two points in space in the statistically stationary state takes the form
\begin{multline}\label{eq:karmanhowarthvorticity}
\nabla_{{\bm{x}}_1} \cdot \langle \bm{u}({\bm{x}}_1,t) \omega({\bm{x}}_1,t) 
\omega({\bm{x}}_2,t) \rangle + \nabla_{{\bm{x}}_2} \cdot \langle \bm{u}({\bm{x}}_2,t)
\omega({\bm{x}}_1,t) \omega({\bm{x}}_2,t) \rangle \\
= \langle \omega({\bm{x}}_1,t)L(-\Delta_{{\bm{x}}_2}) \omega({\bm{x}}_2,t)\rangle + \langle \omega({\bm{x}}_2,t)L(-\Delta_{{\bm{x}}_1}) \omega({\bm{x}}_1,t)\rangle \\
 + \langle \omega({\bm{x}}_1,t)F({\bm{x}}_2,t)\rangle  + \langle \omega({\bm{x}}_2,t)F({\bf
x}_1,t)\rangle
\end{multline} 
which can be derived in a straightforward manner from the evolution equation for vorticity. The transport terms of the left-hand side of the equation are responsible for the transport process in the cascade. They essentially depend on the three-point vorticity correlations $\langle \omega(\bm{x}',t)\omega({\bm{x}}_1,t)\omega({\bm{x}}_2,t) \rangle $ since the velocity field is determined from Biot--Savart's law. They can be re-expressed in terms of the conditional averages by
\begin{equation}
 \langle \bm{u}({\bm{x}}_1,t) \omega({\bm{x}}_1,t) \omega({\bm{x}}_2,t) \rangle =  \int \!
\mathrm{d} \omega_1 \mathrm{d} \omega_2 \ \omega_1 \omega_2 \langle  \bm{u}({\bm{x}}_1,t)
| \omega_1,{\bm{x}}_1 ; \omega_2,{\bm{x}}_2 \rangle f(\omega_1,{\bm{x}}_1;\omega_2,{\bm{x}}_2)
\end{equation}
and analogously for $\langle \bm{u}({\bm{x}}_2,t) \omega({\bm{x}}_2,t) \omega({\bm{x}}_1,t) \rangle$. For the GA, the left-hand side of Eq.\,\eqref{eq:karmanhowarthvorticity} can be explicitly evaluated using Eq.~\eqref{condvelLGA} leading to
\begin{multline}
\int \! \mathrm{d} \omega_1 \mathrm{d} \omega_2 \ \omega_1 \omega_2 
\bigl[\bm{U}({\bm{x}}_1-{\bm{x}}_2)G(\omega_1,\omega_2,|{\bm{x}}_1-{\bm{x}}_2|)
 \cdot \nabla_{{\bm{x}}_1} \\
 + 
 \bm{U}({\bm{x}}_2-{\bm{x}}_1)G(\omega_2,\omega_1,|{\bm{x}}_2-{\bm{x}}_1|) 
\cdot \nabla_{{\bm{x}}_2} \bigr]
f(\omega_1,{\bm{x}}_1;\omega_2,{\bm{x}}_2) \, .
\end{multline}
However, as for homogeneous isotropic statistics the probability distribution depends on $|{\bm{x}}_1-{\bm{x}}_2|$ only, the transport term vanishes since
\begin{equation} 
\bm{U}({\bm{x}}_1-{\bm{x}}_2)= \int \! \mathrm{d}\bm{x}' \, \bm{e}_z \times \frac{{\bm{x}}_1-\bm{x}'}{2\pi|{\bm{x}}_1-\bm{x}'|^2} C(\bm{x}'- {\bm{x}}_2) 
\end{equation}
and the gradient of $f(\omega_1,{\bm{x}}_1;\omega_2,{\bm{x}}_2)$ points in the direction of the vector ${\bm{x}}_1-{\bm{x}}_2$ such that
\begin{equation}
\bm{U}({\bm{x}}_1-{\bm{x}}_2)
\cdot \nabla_{{\bm{x}}_1} f(\omega_1,{\bm{x}}_1;\omega_2,{\bm{x}}_2)=0
\end{equation}
leaving us with
\begin{equation}
  \langle \bm{u}({\bm{x}}_1,t) \omega({\bm{x}}_1,t) \omega({\bm{x}}_2,t) \rangle = \langle \bm{u}({\bm{x}}_2,t) \omega({\bm{x}}_1,t) \omega({\bm{x}}_2,t) \rangle = 0 \, .
\end{equation}
This immediately implies that also the energy transfer term vanishes, which is related to the velocity triple correlation arising in the von~K\'{a}rm\'{a}n--Howarth equation for the velocity. These triple correlations can be related to the velocity-vorticity cross-correlations according to
\begin{multline}
\langle u_{\alpha}({\bm{x}}_1,t) \, u_{\beta}({\bm{x}}_1,t) \, u_{\gamma}({\bm{x}}_2,t) \rangle = \\
\frac{1}{(2\pi)^2}
\int \! \mathrm{d}\bm{x}' \, \mathrm{d}\bm{x}'' \,  \left[\bm{e}_z \times\frac{{\bm{x}}_1-\bm{x}'}{|{\bm{x}}_1-\bm{x}'|^2}\right]_{\beta} \left[\bm{e}_z \times\frac{{\bm{x}}_2-\bm{x}''}{|{\bm{x}}_2-\bm{x}''|^2}\right]_{\gamma} \langle u_{\alpha}({\bm{x}}_1,t) \omega(\bm{x}',t) \omega(\bm{x}'',t) \rangle
\end{multline}
and for $\langle u_{\alpha}({\bm{x}}_2,t) \, u_{\beta}({\bm{x}}_2,t) \, u_{\gamma}({\bm{x}}_1,t) \rangle$ analogously. As the integrand vanishes, we directly obtain
\begin{equation}
  \langle u_{\alpha}({\bm{x}}_1,t) \, u_{\beta}({\bm{x}}_1,t) \, u_{\gamma}({\bm{x}}_2,t) \rangle = 0
\end{equation}
in the GA.

That means, although the linear GA gives good qualitative results, it lacks a central hallmark  of turbulence, the enstrophy and energy transfer across scales. This shows, that while the GA already gives some gross statistical features, non-Gaussian effects have to be taken into account to establish a complete theory.

\section{\label{sec:vorticity_increments}Vorticity Increments}

In this section we leave the linear GA behind and try to find explicit solutions to the PDF equation based on further observations from DNS data. To this end, it is particularly useful to focus on vorticity increments.

The PDF equation for the vorticity increments is derived analogously to the multi-point PDF equation. We start from the evolution equation for the increment
\begin{equation}
\Omega(\bm{x},\bm{r},t)=\omega(\bm{x}+\bm{r},t)-\omega(\bm{x},t)
\end{equation}
which is obtained from the vorticity equation. It reads
\begin{eqnarray}
\frac{\partial }{\partial t} \Omega(\bm{x},\bm{r},t)
+\bm{u}(\bm{x},t)\cdot \nabla_{\bm{x}} \Omega(\bm{x},\bm{r},t)
&+& \bm{v}(\bm{x},\bm{r},t)\cdot \nabla_{\bm{r}} \Omega(\bm{x},\bm{r},t) \nonumber \\
&=& \tilde{L}(-\Delta_{\bm{r}})\Omega(\bm{x},\bm{r},t)+
\tilde{F}(\bm{x},\bm{r},t) \, .
\end{eqnarray}
Thereby, we have defined the forcing increment
\begin{equation}
\tilde F(\bm{x},\bm{r},t)=F(\bm{x}+\bm{r},t)-F(\bm{x},t)
\end{equation}
the velocity increment
\begin{equation}
\bm{v}(\bm{x},\bm{r},t)=\bm{u}(\bm{x}+\bm{r},t)-\bm{u}(\bm{x},t) \, ,
\end{equation}
as well as the operator
\begin{equation}
\tilde{L}(-\Delta_{\bm{r}})\Omega(\bm{x},\bm{r},t)=L(-\Delta_{\bm{r}})\Omega(\bm{x},\bm{r},t)- 
[L(-\Delta_{\bm{r}})\Omega(\bm{x},\bm{r},t)]_{\bm{r}=\bm{0}} \, .
\end{equation}
We now introduce the probability distribution of the vorticity increment statistics
\begin{equation}
H(\Omega,r)=\langle \delta(\Omega-\Omega(\bm{x},\bm{r},t))\rangle \, .
\end{equation}
Due to stationarity, homogeneity, and isotropy this quantity only depends on $\Omega$ and $r$. Furthermore, it is related to the two-point vorticity PDF
$f(\omega_1,\omega_2,r)$ according to   
\begin{equation}
H(\Omega,r)=\int \! \mathrm{d} \omega_1  \mathrm{d}\omega_2 \ \delta(\Omega-(\omega_1-\omega_2)) \, f(\omega_1,\omega_2,r) \, .
\end{equation}
An evolution equation for the PDF $H(\Omega,r)$ is obtained in a similar way as outlined in  Appendix \ref{app:pdf} and reads 
\begin{equation}\label{pdfequation}
\frac{1}{r}\frac{\partial }{\partial r} r V_r(\Omega,r) H(\Omega,r) = -\frac{\partial }{\partial \Omega} \mu(\Omega,r) H(\Omega,r) \, .
\end{equation}
Here, $V_r(\Omega,r)$ is the longitudinal conditional velocity
increment
\begin{equation}
V_r(\Omega,r) = \bigg\langle \frac{\bm{r}}{r}\cdot\bm{v}(\bm{x},\bm{r},t)\bigg|\Omega \bigg\rangle
\end{equation}
and the quantity $\mu(\Omega,r)$ denotes the conditional dissipative term and the conditional forcing:
\begin{equation}\label{eq:def:mue}
\mu(r,\Omega)=\langle \tilde{L}(-\Delta_{\bm{r}})\Omega(\bm{x},\bm{r},t)|\Omega \rangle
+\langle \tilde{F}(\bm{x},\bm{r},t)|\Omega \rangle
\end{equation}
Again, these conditional expectations can be assessed by DNS.

\begin{figure}
\centering
\includegraphics{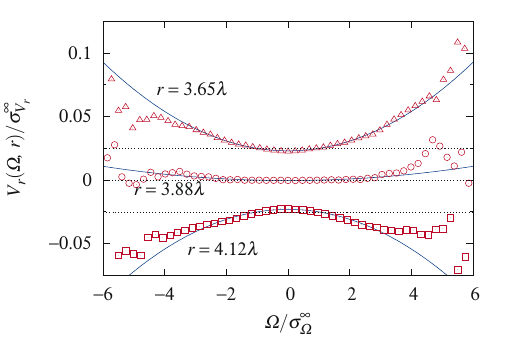}
\caption{\label{some_fits_draft}The numerically estimated term $V_r(\Omega,r)$ for three different distances $r$ together with the quadratic fits according to Eq.\,\eqref{ansatzCond}. For clarity the upper and lower curve have been shifted by $\pm 0.025$; the estimation was based on simulation A.}
\end{figure} 

\begin{figure}
\centering
\includegraphics{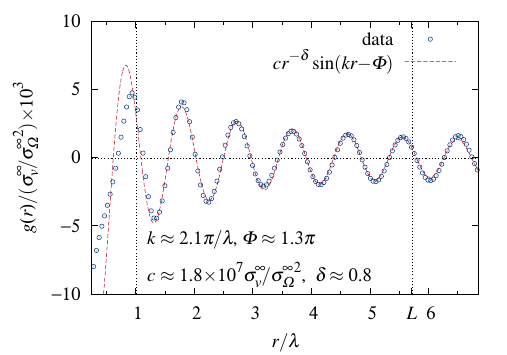}
\caption{\label{grfig}The function $g(r)$ and the fit according to
Eq.\,\eqref{gr}. Scale is measured in terms of the Taylor scale $\lambda$. The integral scale is denoted by $L$. We used the constants $\sigma_v^\infty = (2\langle \bm{u}^2\rangle)^{1/2}$ and $\sigma_\Omega^\infty = (2\langle \omega^2 \rangle)^{1/2}$ respectively to normalize our results.
}
\end{figure}

\begin{figure}
\centering
\includegraphics{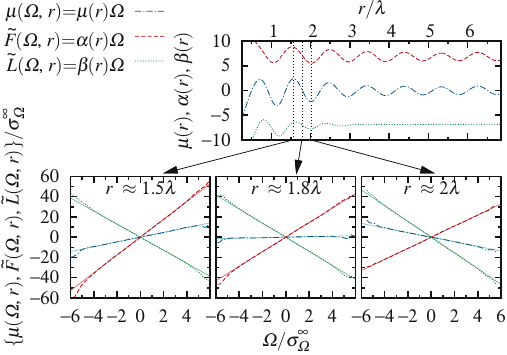}
\caption{
\label{mue} The contributions to the function $\mu(r,\Omega)$ [see Eq.\,\eqref{eq:def:mue}], fitted by $\mu(\Omega,r)=\mu(r) \Omega$, due to forcing, $\tilde{F}(\Omega,r) = \langle F (\bm{x},\bm{r}, t) \vert \Omega \rangle = \alpha(r)\Omega$, as well as (combined) viscosity and friction, $\tilde{L}(\Omega, r) = \langle L(-\Delta_r)\Omega (\bm{x},\bm{r}, t) \vert \Omega \rangle = \beta(r)\Omega$. The upper panel shows the evolution of the fit parameters $\mu(r)$, $\alpha(r)$, and $\beta(r)$ as a function of the scale $r$. The three graphs in the bottom show data obtained from simulation~A as well as our linear fits (solid lines) for three constant $r$.
}

\end{figure}

The conditional longitudinal velocity increment $V_r(\Omega,r)$ has to be an even function of $\Omega$, 
furthermore its average has to vanish:
\begin{eqnarray}\label{cond}
V_r(\Omega,r) &=& V_r(-\Omega,r) \nonumber \\
\int \! \mathrm{d}\Omega \ H(\Omega,r) V_r(\Omega,r)&=& 0 \, .
\end{eqnarray}
We have determined the conditional velocity increment $V_r(\Omega,r)$ as well as the conditionally averaged terms  $\mu(\Omega,r)$ of the right-hand side of Eq.~\eqref{pdfequation} from DNS  (see Figures~\ref{some_fits_draft} and~\ref{mue}). It turns out that for not too small values of $r$ it is possible to approximate the function using 
the following quadratic ansatz in $\Omega^2$,
\begin{equation}\label{ansatzCond}
V_r(\Omega,r) = g(r) [\Omega^2-\langle \Omega(r,t)^2 \rangle] \, ,
\end{equation}
which is consistent with the conditions~\eqref{cond} and can be viewed as a Taylor expansion in $\Omega$. From Figure~\ref{grfig} it becomes clear that the function $g(r)$ is an oscillating function of $r$ which  in the inertial range can be well approximated by
\begin{equation}\label{gr}
g(r)= c r^{-\delta} \sin \left(k r -\Phi \right) \, .
\end{equation}
The constant $k=2.1\pi / \lambda$ is related to the width of the forcing function of the turbulence. For the exponent we find $\delta \approx 0.8$. Amplitude and phase turn out to be $c=1.8\times 10^7 \sigma_{\nu}^{\infty}/ \sigma_{\Omega}^{\infty 2}$ and $\Phi=1.3\pi$.
  
The function $\mu(\Omega,r)$ consists of three contributions, namely forcing, dissipation, and friction. For the central core of the PDFs, all contribute with a linear dependency in $\Omega$ to $\mu(\Omega,r)$ as can be seen from our DNS data in Figure~\ref{mue}. This leads us to the ansatz \begin{equation}\label{ansatzCond2}
  \mu(\Omega,r) = \mu(r) \Omega \, .
\end{equation}

\begin{figure}
\centering
\includegraphics{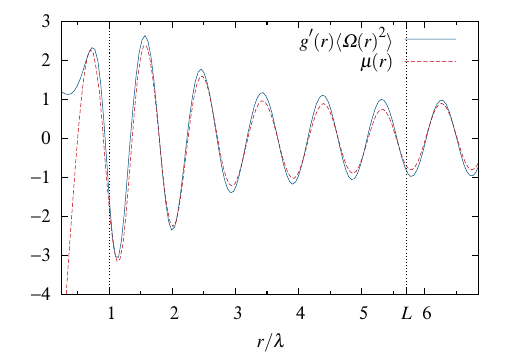}
\caption{\label{checkf}The function $\mu(r)$ for the contributions due to forcing, viscosity and friction compared with the relation $\mu(r)=\langle \Omega(r)^2 \rangle g'(r)$. Scale is measured in terms of the Taylor scale. The shown data has been calculated from the results of our numerical simulation~A.}
\end{figure}

It is interesting to notice that we can find a solution for the PDF from the kinetic equation for values of the scale, for which the function $g(r)$ vanishes. In that case the characteristic equation degenerates to the differential equation 
\begin{equation}
g'(r) [\Omega^2-\langle \Omega^2(r,t) \rangle] H(\Omega,r)=
-\frac{\partial }{\partial \Omega} \mu(r) \Omega H(\Omega,r)
\end{equation}
with the explicit Gaussian solution
\begin{equation}\label{gauss}
H(\Omega,r)=\frac{\mathrm{e}^{-\frac{\Omega^2}{2 \langle \Omega(r,t)^2\rangle }}}
{\sqrt{2\pi \langle \Omega(r,t)^2\rangle} } \, .
\end{equation}
provided the following non-trivial relationship between conditional velocity field, conditional dissipation and forcing, and the correlation function $\langle \Omega(r,t)^2 \rangle$ holds:
\begin{equation}
 \mu(r) = g'(r) \langle \Omega(r,t)^2 \rangle
\end{equation}
A check of this equality is presented in Fig.\,\ref{checkf}. We want to point out that, strictly, the result is only valid for values of $r$, for which $g(r)=0$, i.e. at the minima and maxima of the moment. In between these points, the PDF can be determined perturbatively. Comparing this explicit Gaussian solution to the DNS data in Figure~\ref{pdf}, our approximate solution provides a good description for not too large increments. As the Gaussian solution depends on the correlation function, it also exhibits the typical slight oscillations.
As a last remark, we remind the reader that the kinetic equation determining
the vorticity increment PDF is a nonlinear equation for $H(\Omega,r)$, since the
conditional longitudinal velocity increment $V_r(\Omega,r)$ contains the expectation value $\langle \Omega(r,t)^2 \rangle$.

\section{Summary and Conclusions}

In summary, we have presented a statistical description of the vorticity statistics in the inverse cascade regime of two-dimensional turbulence in terms of the PDF equations of the LMN hierarchy. The unclosed terms arising in this framework as unknown conditional averages have been evaluated from DNS data and compared to an analytically tractable GA. While this comparison shows good qualitative agreement, this approximation is not able to explain the inverse energy cascade. As a result, the energy flux across scales can be regarded as non-perturbative.

The interpretation of the multi-point PDF equation in terms of the method of characteristics has led to a semi-Lagrangian statistical description. In particular, we were able to derive an effective point vortex description, where the vortex profiles need to be modified due to the average effects of turbulence.  
The GA here has turned out to be particularly useful to explicitly calculate a possible modification of a point vortex profile. The main result here is that the effective azimuthal dynamics is a two-vortex motion, however with a screened velocity field. The effective relative motion, which is absent for bare point vortices as well as for the vortex motion in the linear GA, is the fingerprint of the inverse cascade. It determines the energy and enstrophy fluxes. The relative motion can be attracting or repelling, depending on the distance and the instantaneous vorticities. In Ref.\,\onlinecite{friedrich10cp} it has been argued that the energy transport in the inverse cascade may be viewed as a genuinely nonlinear transport process similar to the 
dynamics of a ratchet. A ratchet process, generally speaking, transforms worthless noise into directed motion. This is essentially what happens in the inverse cascade. 

As an outlook we reiterate that the signature of the inverse cascade as evidenced in the two-point vorticity statistics is encoded in the conditional velocity increment investigated in this article, leading to an attraction and repulsion of two effective quasi-vortices. This effective interaction seems to be a non-perturbative effect, and it remains a challenge to derive this relative motion directly from the stochastic Navier-Stokes dynamics. Recently, a vortex model has been formulated, which takes into account a deformation of vortices, i.e. the so-called vortex-thinning, due to shear generated by distant vortices \cite{Friedrich2011arx}. In this model like-signed vortices cluster and it is possible to determine the relative motion of two vortices, which has been identified in the present work, directly from the model equation.In this respect, clustering of vorticity patches with like-signed circulation is the mechanism underlying the inverse cascade. However, due to the small-scale forcing, the emerged larger-scale vortices are permanently destroyed avoiding the formation of big vortices of the size of the system. This view is consistent with the experimental results of Paret and Tabeling \cite{paret97prl}, who identified a kind of agregation process of like-signed vorticity patches. Furthermore, it supports ideas about the relation between the inverse cascade and a kind of Bose condensation mechanism, as has been discussed by Smith and Yakhot \cite{Smith93prl}, Chertkov et al. \cite{Chertkov2007prl}. We hope that the present analysis of two-point vorticity statistics will help to shed light on this collective phenomenon.     

\section*{Acknowledgments}

We note with sadness that Professor Rudolf Friedrich unexpectedly passed away on August 16th 2012. He has been a dedicated teacher, caring mentor and inspiring scientist. He will be missed by those who knew him.

\appendix

\vspace{0.25cm}

\section{\label{app:pdf}Kinetic Equations} 
For the sake of completeness, we present the derivation of the hierarchy of evolution equations.  
Here we confine our discussion to the simpler case of a one-point vorticity PDF but the extension to the $N$-point case is straightforward. As starting point we define the fine-grained one-point vorticity distribution 
\begin{equation}
\tilde f(\omega_1,{\bm{x}}_1,t) = \delta(\omega_1-\omega({\bm{x}}_1,t)) \, .
\end{equation}
The evolution equation for this quantity is obtained by differentiation with respect to time:
\begin{eqnarray}
\frac{\partial }{\partial t} \tilde f(\omega_1,{\bm{x}}_1,t) = -\left[\frac{\partial }{\partial t}\omega({\bm{x}}_1,t)\right]\frac{\partial }{\partial \omega_1} \tilde f(\omega_1,{\bm{x}}_1,t)=-\frac{\partial }{\partial \omega_1} \left[\frac{\partial }{\partial t}\omega({\bm{x}}_1,t)\right] \tilde f(\omega_1,{\bm{x}}_1,t)
\end{eqnarray}
Taking into account the vorticity equation we obtain
\begin{equation}
\frac{\partial }{\partial t}\tilde f(\omega_1,{\bm{x}}_1,t) + \bm{u}({\bm{x}}_1,t)\cdot \nabla_{{\bm{x}}_1}\tilde f(\omega_1,{\bm{x}}_1,t)= -\frac{\partial }{\partial \omega_1} \big[L(-\Delta_{{\bm{x}}_1}) \omega({\bm{x}}_1,t) + F({\bm{x}}_1,t)\big] \tilde f(\omega_1,{\bm{x}}_1,t)  \, .
\end{equation}
Thereby, we have used the fact that the velocity field is incompressible,
\begin{align}
 \bm{u}({\bm{x}}_1,t)\cdot \nabla_{{\bm{x}}_1} \tilde f(\omega_1,{\bm{x}}_1,t) 
 &= -\left[\bm{u}({\bm{x}}_1,t)\cdot \nabla_{{\bm{x}}_1} \omega({\bm{x}}_1,t)\right] \frac{\partial }{\partial \omega_1} \tilde f(\omega_1,{\bm{x}}_1,t)  \\
 &= -\frac{\partial }{\partial \omega_1} \left[\bm{u}({\bm{x}}_1,t)\cdot \nabla_{{\bm{x}}_1} \omega({\bm{x}}_1,t)\right] \tilde f(\omega_1,{\bm{x}}_1,t) \, .
\end{align}
We now proceed to derive the evolution equations for the ensemble probability distribution. Averaging yields
\begin{align}
\frac{\partial }{\partial t} f(\omega_1,{\bm{x}}_1,t) + \nabla_{\bm{x}_1} \cdot \langle \bm{u}({\bm{x}}_1,t) \tilde f(\omega_1,{\bm{x}}_1,t) \rangle 
&= \nonumber \\ -\frac{\partial }{\partial \omega_1} \langle L(-\Delta_{{\bm{x}}_1}) \omega({\bm{x}}_1,t) \tilde f(\omega_1,{\bm{x}}_1,t) \rangle \, &-\frac{\partial }{\partial \omega_1} \langle F({\bm{x}}_1,t) \tilde f(\omega_1,{\bm{x}}_1,t)\rangle  .
\end{align}
The unclosed expectation values, e.g. $ \langle \bm{u}({\bm{x}}_1,t) \tilde f(\omega_1,{\bm{x}}_1,t) \rangle$, can be expressed in terms of the conditional expectation
\begin{equation} \label{eq:unclosedCond}
\langle \bm{u}({\bm{x}}_1,t) \tilde f(\omega_1,{\bm{x}}_1,t)\rangle=\langle \bm{u}({\bm{x}}_1,t) | \omega_1, {\bm{x}}_1 \rangle f(\omega_1,{\bm{x}}_1,t) \, ,
\end{equation}
Using this relation for all expectation values leads to an evolution equation for the one-point vorticity PDF of the form
\begin{align}
 \frac{\partial }{\partial t} f(\omega_1, {\bm{x}}_1,t ) + \nabla_{{\bm{x}}_1} \cdot \big[\langle \bm{u}({\bm{x}}_1,t)|\omega_1, {\bm{x}}_1 \rangle f(\omega_1, {\bm{x}}_1,t)\big] &= \nonumber \\ -\frac{\partial }{\partial \omega_1}  \big[ \langle L(-\Delta_{{\bm{x}}_1}) \omega({\bf
x}_1,t)|\omega_1, {\bm{x}}_1 \rangle &+  \langle F({\bm{x}}_1,t)|\omega_1, {\bm{x}}_1 \rangle \big]
f(\omega_1, {\bm{x}}_1,t) \, .
\end{align}
where all unclosed terms are expressed as conditional averages. Alternatively one could proceed by using Biot--Savart's law~\eqref{eq:BiotSavart} to rewrite the term~\eqref{eq:unclosedCond} in  in terms of the two-point probability distribution \cite{lundgren67pf}
\begin{align}
\langle \bm{u}({\bm{x}}_1,t) \tilde f(\omega_1,{\bm{x}}_1,t)\rangle &=
\int \! \mathrm{d}{\bm{x}}_2 \int \! \mathrm{d}\omega_2 \ \bm{e}_z \times
\frac{{\bm{x}}_1-{\bm{x}}_2}{2\pi |{\bm{x}}_1-{\bm{x}}_2|^2} \omega_2
f(\omega_1,{\bm{x}}_1;\omega_2,{\bm{x}}_2,t)\,.
\end{align} 
In a similar way, the dissipative term can be expressed as
\begin{equation}
\langle  L(-\Delta_{{\bm x}_1}) \omega({\bm{x}}_1,t) \tilde f(\omega_1,{\bm{x}}_1,t) \rangle=
 \int \! \mathrm{d}{\bm{x}}_2 \int \! \mathrm{d} \omega_2 \ \delta({\bm{x}}_1-{\bm{x}}_2)
\omega_2 L(-\Delta_{{\bm x}_2})
f(\omega_1,{\bm{x}}_1;\omega_2,{\bm{x}}_2,t) \, .
\end{equation} 
In this formulation we can explicitly see that due to the unclosed expectation values the evolution of the one-point PDF couples the the two-point vorticity PDF.

\section{\label{app:lga}Conditional Expectations: Gaussian Approximation} 

It is straightforward to evaluate the expectation necessary to calculate the conditional velocity field based on the assumption of Gaussian statistics using the Fourier representation of the distribution \cite{friedrich10cp}. By using the conventional definition of the characteristic function we can write the $N+1$-point distribution as
\begin{equation}
f(\omega',\bm{x}',\lbrace \omega_l,{\bm{x}}_l \rbrace)= \frac{1}{(2\pi)^{N+1}} \int \! \mathrm{d} \alpha' \, \Pi_l  \, \mathrm{d}\alpha_l \mathrm{e}^{-\mathrm{i} (\alpha' \omega'+\sum_l \alpha_l \omega_l)} \mathrm{e}^{-W(\alpha',\bm{x}',\lbrace \alpha_l,{\bm{x}}_l \rbrace)} \, .
\end{equation}
The quantity $W(\alpha',\bm{x}',\lbrace \alpha_l,{\bm{x}}_l \rbrace)$ is the cumulant-generating function of the multi-point vorticity statistics. The conditional expectation of vorticity is then obtained by the following calculation:
\begin{multline}
\int \! \mathrm{d} \omega' \, \omega' f(\omega',\bm{x}';\lbrace \omega_l,{\bm{x}}_l \rbrace)\\
 \begin{aligned}[b]
  &=\frac{1}{(2\pi)^{N+1}}  \int \! \mathrm{d} \omega' \, \omega' \int \! \mathrm{d} \alpha' \, \Pi_l  \, \mathrm{d}\alpha_l \mathrm{e}^{-\mathrm{i} (\alpha' \omega'+\sum_l \alpha_l \omega_l)} \mathrm{e}^{-W(\alpha',\bm{x}',\lbrace \alpha_l,{\bm{x}}_l \rbrace)}
 \\
  &= \frac{1}{(2\pi)^{N+1}}  \int \! \mathrm{d} \omega' \int \! \mathrm{d} \alpha' \, \Pi_l  \, \mathrm{d}\alpha_l  \left[-\frac{1}{\mathrm{i}} \frac{\partial }{\partial \alpha'} \mathrm{e}^{-\mathrm{i} (\alpha' \omega'+\sum_l \alpha_l \omega_l)} \right] \mathrm{e}^{-W(\alpha',\bm{x}',\lbrace \alpha_l,{\bm{x}}_l \rbrace)}
 \\
  &= - \frac{1}{(2\pi)^{N+1}} \int \! \mathrm{d} \omega' \int \! \mathrm{d} \alpha' \, \Pi_l  \, \mathrm{d}\alpha_l \mathrm{e}^{-\mathrm{i} (\alpha' \omega'+\sum_l \alpha_l \omega_l)}  \left[\frac{1}{\mathrm{i}} \frac{\partial }{\partial \alpha'} W(\alpha',\bm{x}',\lbrace \alpha_l,{\bm{x}}_l \rbrace) \right] \mathrm{e}^{-W(\alpha',\bm{x}',\lbrace \alpha_l,{\bm{x}}_l \rbrace)}
 \\
  &= - \frac{1}{(2\pi)^{N}} \int \! \mathrm{d} \alpha' \Pi_l  \, \mathrm{d}\alpha_l \mathrm{e}^{-\mathrm{i}\sum_l \alpha_l \omega_l}  \left[\frac{1}{\mathrm{i}} \frac{\partial }{\partial \alpha'} W(\alpha',\bm{x}',\lbrace \alpha_l,{\bm{x}}_l \rbrace) \right]
    \mathrm{e}^{-W(\alpha',\bm{x}',\lbrace \alpha_l,{\bm{x}}_l \rbrace)}  \frac{1}{2\pi}\int \! \mathrm{d} \omega' \, \mathrm{e}^{-\mathrm{i} \alpha' \omega'} \\
  &= -\frac{1}{(2\pi)^{N}} \int \! \Pi_l  \, \mathrm{d}\alpha_l \mathrm{e}^{-\mathrm{i}\sum_l \alpha_l \omega_l}  \left[ \frac{1}{\mathrm{i}} \frac{\partial }{\partial \alpha'} W(\alpha',\bm{x}',\lbrace \alpha_l,{\bm{x}}_l \rbrace) \right]_{\alpha' = 0} \mathrm{e}^{-W(\lbrace \alpha_l,{\bm{x}}_l \rbrace)}  \\
  &= - \left[\frac{1}{\mathrm{i}} \frac{\partial }{\partial \alpha'} W\left(\alpha',\bm{x}',\left \lbrace  -\frac{1}{\mathrm{i}} \frac{\partial }{\partial \omega_l} ,{\bm{x}}_l \right\rbrace\right)\right]_{\alpha'=0}  \frac{1}{(2\pi)^{N}} \int \Pi_l \mathrm{d}\alpha_l \mathrm{e}^{\mathrm{i} \sum_l \alpha_l \omega_l} \mathrm{e}^{-W(\lbrace \alpha_l,{\bm{x}}_l \rbrace)} \\
  &=-\left[\frac{1}{\mathrm{i}} \frac{\partial }{\partial \alpha'} W\left(\alpha',\bm{x}',\left\lbrace -\frac{1}{\mathrm{i}} \frac{\partial }{\partial \omega_l} ,{\bm{x}}_l \right\rbrace\right)\right]_{\alpha'=0} f(\lbrace \omega_l,{\bm{x}}_l \rbrace) \, .
 \end{aligned}
\end{multline}
This is an exact relationship involving the generating function $W(\alpha',\bm{x}',\lbrace \alpha_l ,{\bm{x}}_l \rbrace)$ of the $(N+1)$-point cumulants of vorticity. The Gaussian $N$-point distribution is defined as
\begin{equation}
f(\lbrace \omega_l,{\bm{x}}_l \rbrace)= (2\pi)^{-\frac{N}{2}} \mbox{det}(\mathrm{C})^{-\frac{1}{2}} \mathrm{e}^{-\frac{1}{2} \sum_{k,m=1}^N \omega_k C^{-1} ({\bm{x}}_k-{\bm{x}}_m) \omega_m } \, ,
\end{equation}
where $\mathrm{C}$ denotes the $N \times N$-matrix containing the two-point correlation functions $C(\bm{x}_i-\bm{x}_j)$, and $C^{-1}(\bm{x}_i-\bm{x}_j)$ denotes the $(i,j)$-th element of the inverse of $\mathrm{C}$. Together with the Gaussian $N+1$-point cumulant-generating function  
\begin{equation}
W(\alpha',\bm{x}',\lbrace \alpha_l,{\bm{x}}_l \rbrace) = \frac{\alpha'^2}{2} C(0) +\sum_{km} \frac{\alpha_i\alpha_j}{2} C({\bm{x}}_k-{\bm{x}}_m) +\sum_k \alpha' \alpha_k C(\bm{x}'-{\bm{x}}_k)
\end{equation}
we obtain for the expectation
\begin{align}
\int \! \mathrm{d} \omega' \, \omega' f(\omega',\bm{x}';\lbrace \omega_l,{\bm{x}}_l\rbrace) &= -\left[\frac{1}{\mathrm{i}} \frac{\partial }{\partial \alpha'} W\left(\alpha',\bm{x}',\left\lbrace -\frac{1}{\mathrm{i}} \frac{\partial }{\partial \omega_l} ,{\bm{x}}_l \right\rbrace\right)\right]_{\alpha'=0} f(\lbrace \omega_l,{\bm{x}}_l \rbrace) \nonumber \\
&= \sum\limits_k C(\bm{x}'-{\bm{x}}_k) \frac{\partial }{\partial \omega_k} f(\lbrace \omega_l,{\bm{x}}_l \rbrace) \nonumber \\
&= \sum\limits_{km} C(\bm{x}'-{\bm{x}}_k) C^{-1}({\bm{x}}_k-{\bm{x}}_m) \omega_m f(\lbrace \omega_l, {\bm{x}}_l \rbrace) \nonumber \\
&= \langle  \omega',\bm{x}'|\lbrace \omega_l,{\bm{x}}_l\rbrace \rangle f(\lbrace \omega_l, {\bm{x}}_l \rbrace) \, .
\end{align}
This expression for the conditional vorticity can be used to calculate the conditional velocity~\eqref{condvel} as well as the conditional dissipation~\eqref{conddiss} in the Gaussian Approximation. We note that in the Gaussian Approximation the conditional vorticity is a linear function of the sample-space vorticities.

\end{document}